\documentclass[%
 reprint,
 amsmath,amssymb,
 aip,
]{revtex4-1}

\usepackage[english]{babel}
\usepackage{graphicx}% Include figure files
\usepackage{dcolumn}% Align table columns on decimal point
\usepackage{bm}% bold math
\begin{document}

\preprint{APS/123-QED}

\title{Towards 3D Magnetic Force Microscopy}

\author{Jori F. Schmidt}
\affiliation{TU Dresden, Institute of Applied Physics, Nöthnitzer Strasse 61, 01187 Dresden, Germany}

\author{Lukas M. Eng}%
\affiliation{TU Dresden, Institute of Applied Physics, Nöthnitzer Strasse 61, 01187 Dresden, Germany}
\affiliation{ct.qmat: Dresden-Würzburg Cluster of Excellence—EXC 2147, TU Dresden, 01062 Dresden, Germany}
\email{lukas.eng@tu-dresden.de}

\author{Samuel D. Seddon}
\affiliation{TU Dresden,  Institute of Applied Physics, Nöthnitzer Strasse 61, 01187 Dresden, Germany}
\email{samuel.seddon@tu-dresden.de}

\date{\today}% It is always \today, today,
             %  but any date may be explicitly specified

\begin{abstract}
Magnetic force microscopy (MFM) is long established as a powerful tool for probing the local manifestation of magnetic nanostructures across a range of temperatures and applied stimuli. A major drawback of the technique, however, is that the detection of stray fields emanating from a samples surface rely on a uniaxial vertical cantilever oscillation, and thus are only sensitive to vertically oriented stray field components. The last two decades have shown an ever-increasing literature fascination for exotic topological windings where particular attention to in-plane magnetic moment rotation is highly valuable when identifying and understanding such systems. Here we present a new method of detecting in-plane magnetic stray field components, by utilizing a home made split-electrode excitation piezo that allows the simultaneous excitation of a cantilever at its fundamental flexural and torsional modes. This allows for the joint acquisition of traditional vertical mode (V-MFM) images and a lateral MFM (L-MFM) where the tip-cantilever system is only sensitive to stray fields acting perpendicular to the torsional axis of the cantilever. 
\end{abstract}

%\keywords{Suggested keywords}%Use showkeys class option if keyword
                              %display desired
\maketitle

%\tableofcontents

\begin{figure*}[ht!]
\includegraphics[]{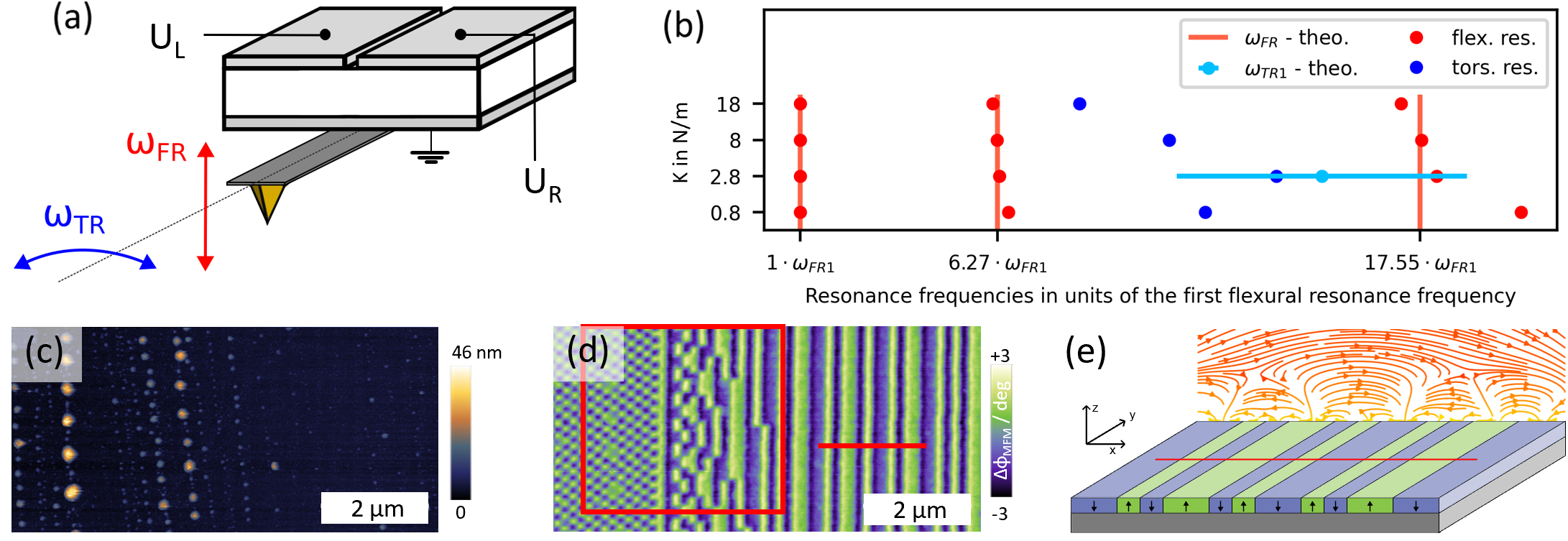}
\caption{\label{fig:intro} (a) shows a sketch of the custom cantilever holder with two separate driving electrodes. The flexural and torsional resonant frequencies were measured for different AFM cantilevers and are plotted in (b), each normalized to the first flexural resonant frequency. (c) is a topography scan of the hard disk while (d) shows a conventional V-MFM image of the same area. Along the red line, linescans were performed, the reconstructed magnetization structure in that region is sketched in (e), together with the simulated stray field.}
\end{figure*}
\section{\label{sec:Introduction}Introduction}
Non-contact Scanning Force Microscopy (nc-SFM) typically utilizes a purely flexural cantilever excitation in order to attain real space resolved information about a samples surface. Functionalization of an SFM tip, such as the addition of a magnetic coating, allows for the detection of magnetic structures on the samples surface as well. 
Traditional, vertically excited magnetic force microscopy (V-MFM) is sensitive to out-of-plane components of the magnetic stray field above a sample \cite{hartmann_theory_1990, kazakova_frontiers_2019}. The lack of in-plane resolution, however, means that V-MFM provides only a limited visualization of the magnetic structures present. Frequently more relevant in recent years, with the advent of exotic chiral spin windings such as skyrmions \cite{milde_surface_2019,milde_unwinding_2013,zhang_imaging_2016,zhang_multidomain_2016}, antiskyrmions \cite{winter_antiskyrmions_2022, sukhanov_anisotropic_2020}, and more \cite{seddon_real-space_2021}, access to sensitivity of more than one stray field vectorial direction could vastly increase the nanoscale understanding and eventual control of these functionalized materials \cite{fert_magnetic_2017}. 

In the past, attempts at multiple in-plane component sensitive measurements have been made to achieve this by performing V-MFM with a strongly tilted tip \cite{abraham_measurement_1988} or by exciting an in-plane motion of a magnetic monopole tip by using a specially designed cantilever, oscillating at its second flexural mode \cite{muhl_magnetic_2012}. Alternatively, changing the magnetization direction of the V-MFM tip to be aligned parallel to the sample surface has also been used \cite{rice_observation_1999}. The latter method requires specially designed tips \cite{folks_perforated_2000}, as normal V-MFM tips do not reliably retain an in-plane magnetization due to the strong shape anisotropy introduced by the shape of the tip itself. Another draw back of these collected attempts were that whilst successful in changing the direction of stray field sensitivity, they were unable to also maintain standard V-MFM sensitivity simultaneously. 

An alternative approach is pursued in this work, by instead changing the tip oscillation direction to be not purely out-of-plane (with flexural mode excitations) as in traditional non-contact SFM, but instead a torsional, in-plane mode, parallel to the surface of the sample. This is achieved by exciting the torsional resonant mode of a conventional V-MFM cantilever, providing a lateral magnetic force microscopy signal (L-MFM). Such torsional methods have been  previously performed to for example measure frictional properties \cite{gnecco_friction_2003, pfeiffer_lateral-force_2002} or when studying ferroelectrics \cite{eng_ferroelectric_1997}. In-plane tip excitations have also been used to improve topographic imaging of samples that have more challenging structures \cite{zhang_development_2022}. This work will first outline the method and validity of torsional cantilever excitation before demonstrating theoretical and experimental insights into the sensitivity to in-plane magnetic stray field contributions.

\section{\label{sec:Experimental}Experimental methods}
In order to excite both the flexural and the torsional mode of the cantilever simultaneously, a custom tip holder was developed. The tip holder consists of two parallel driving piezos that can be controlled individually, sharing a common ground electrode, as sketched in Fig.~\ref{fig:intro}(a). 
Topography and normal V-MFM was recorded using a commercial amplitude modulation AFM (NX10 from Park Systems). Tips used in this study for magnetic imaging were commercially available PPP-MFMR cantilevers from Nanosensors. On one piezos the usual output of the AFM controller was connected, and consequently resonated at the first flexural resonant frequency ($\sim70$ kHz). The other piezo was connected to the output of an external lock-in amplifier (Zurich Instruments UHF) and driven at the first torsional resonant frequency of the cantilever ($1V$ driving voltage).  The motion of the tip therefore can be thought of as a superposition of both of these two modes, one flexural, and one torsional. As a quadrant photo-diode, the built-in photo-detector of the AFM does not only output the vertical deflection, corresponding to the flexural bending mode normally used for AFM, but also a torsional deflection. In addition to the usual V-MFM phase, the phase between the torsional deflection of the cantilever and the torsional driving signal was measured on the external lock-in and exported as the lateral V-MFM (L-MFM) phase, corresponding to the in-plane magnetic stray field and us used for imaging (pink color scale). Vertical tip oscillations of 25 nm were used, and scanning performed in a dual-pass regime, lift height 15 nm. \\

The V-MFM cantilevers used have a corresponding spring constant of $2.8 ~$N/m. To explore the relative positions between the torsional and flexural resonant frequencies for other AFM cantilevers, a variety of Rocky Mountain cantilevers were chosen with spring constants approximately between 0.8 and 18 N/m, to confirm the reliable operation of the dual frequency tip holder. The values are summarized in Fig.~\ref{fig:intro}(b), where the resonant frequencies of each cantilever are normalized with respect to the first flexural resonant frequency $\omega_{FR1}$.\\

As seen, the torsional frequencies are found to lie always within the second and third flexural resonance mode. To confirm the expected position of the first torsional mode, the theoretical torsional resonant frequency was calculated \cite{augustyn_possibility_2016} and is in good agreement with the experiment for the PPPMFM cantilever. The theoretical torsional frequency could not be calculated for the other tips due to insufficient information about the cantilever geometry.

The sample investigated in this work was a magnetic hard disk, with an array of positive and negative out-of-plane (normal to the surface) magnetizations, a conventional V-MFM image of which is shown in Fig.~\ref{fig:intro}(d). Regions of up and down magnetization are used to encode the data bits. An area consisting of long stripe domains was chosen for imaging, as their simple structure [a schematic of which can be found in Fig.~\ref{fig:intro}(e)], makes it possible to easily model the expected V-MFM and L-MFM response, and interpret the results. 

Simulations of the expected V-MFM and L-MFM response for the hard-disk sample were performed in Python using the Magpylib \cite{ortner_magpylib_2020} library. The library provides the analytical solution for the magnetic stray field of a cuboid, multiple of which can be pieced together to model the hard disk. 

%The linescans are performed with different orientations of the in-plane oscillation direction: first with the in-plane oscillation oriented along the x-direction (orthogonal to the domains) and then with the in-plane oscillation oriented along the y-direction (parallel to the domains). The rotation of the in-plane oscillation direction is achieved by rotating the sample.

\section{\label{sec:theory}Theoretical considerations}
\begin{figure*}[ht]
\includegraphics{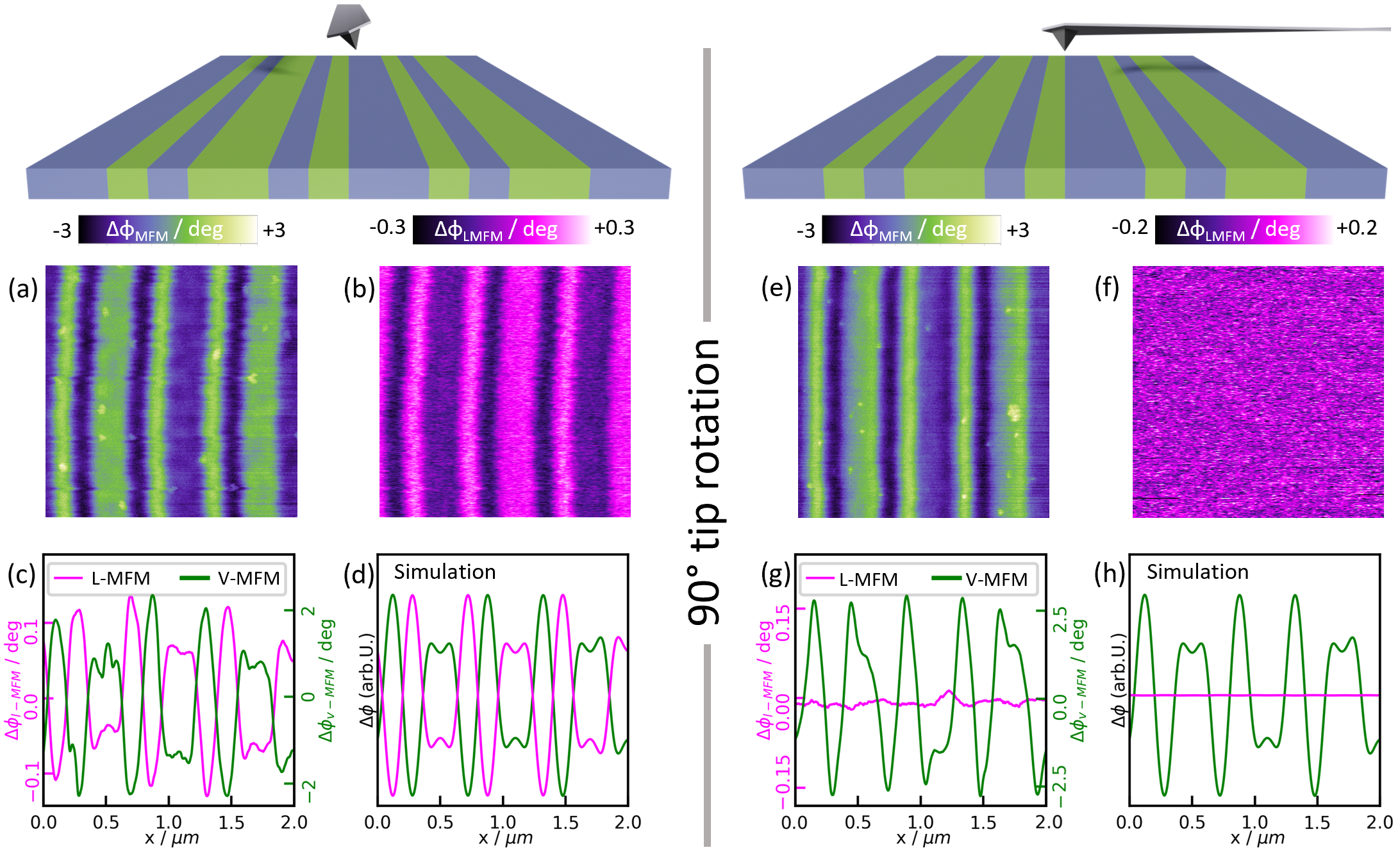}
\caption{\label{fig:linescans} V-MFM and L-MFM linescans for two different in-plane oscillation directions, in (a-d) the in-plane oscillation of the tip is orthogonal to the long stripe domains, in (e-g) it is parallel. (a) shows the V-MFM image, (b) the L-MFM image, an averaged linescan over the same stripe domains is plotted in (c). (d) shows the simulated V-MFM and  L-MFM response for this situation. (e) and (f) show V-MFM and L-MFM images respectively, with the in-plane oscillation of the tip parallel to the stripe domains, averaged linescans are shown in (g) and modeled V-MFM and L-MFM response in (h).}
\end{figure*}

The theoretical interaction between tip and stray field in conventional V-MFM is well established, however here it will be extended for considerations of an in-plane oscillation as well. In the simplest model for the V-MFM response, the tip is modeled as a magnetic point dipole, with its magnetic moment $\vec{m}=m\vec{e_z}$ oriented along the symmetry axis of the tip (the z-direction). The force acting on such a dipole from the samples stray field $\vec{H}$ is then given by
\begin{equation}
    \vec{F} = \mu_0 m  \frac{\partial}{\partial z} \vec{H},
\end{equation}
where $\mu_0$ is the vacuum permeability. For the case of conventional V-MFM, where the tip is oscillating along the z-direction, the difference in phase from free resonance and when a (magnetic) force is present is then commonly approximated by \cite{garcia_amplitude_2010}:
\begin{equation}\label{eq:MFMPhase}
    \Delta\phi_{\text{V-MFM}} \approx \frac{Q}{k} \frac{\partial F_z}{\partial z} = \mu_0 m \frac{Q}{k} \frac{\partial^2}{\partial z^2} H_z,
\end{equation}
where $Q$ and $k$ are the quality factor and spring constant of the flexural oscillation, respectively. When repeating the calculation from \cite{garcia_amplitude_2010} for a tip oscillating along the x-direction (with the y-direction being the long axis of the cantilever), the expected change in phase is then:
\begin{equation}\label{eq:LMFMPhase}
\begin{split}
    \Delta\phi_{\text{L-MFM}} \approx \frac{Q}{k} \frac{\partial F_x}{\partial x} &= \mu_0 m \frac{Q}{k} \frac{\partial^2}{\partial x \partial z} H_x \\
    &= \mu_0 m \frac{Q}{k} \frac{\partial^2}{\partial x^2} H_z.
\end{split}
\end{equation}
The last equality holds since $\vec{\nabla} \times \vec{H} = 0$. Note that $Q$ and $k$ now refer to the torsional oscillation. The L-MFM signal is therefore expected to be sensitive to changes of the magnetic stray field along the direction of the in-plane tip oscillation, i.e. stray fields acting perpendicular to both the torsional axis and flexural oscillation direction. \\

\section{\label{sec:Results}Results and discussion}

\subsubsection{V-MFM, L-MFM and rotation of the torsional axis}
As established in the ``theoretical considerations'' section, a phase shift in the L-MFM signal should only be expected whenever there is a stray field component along the torsional oscillation direction. In the case of the stripe domains from the hard disk sample, the stray field is symmetric along the stripes, meaning no signal is expected, however, when crossing a domain wall, a significant in-plane field component arises, as sketched in Fig.~\ref{fig:intro}(e). Whilst in this experimental set up, the torsional axis is fixed by the AFM itself, the relative position of the torsional axis with respect to the sample can be rotated by rotating the sample relative to the cantilever. As such, Fig.~\ref{fig:linescans} is split into two halves, with the relative cantilever (and thus the torsional axis) indicated above (a-d) and (e-h), respectively. 

Fig.~\ref{fig:linescans}(a) and (b) show V-MFM and L-MFM images, respectively, of the stripe domains on the magnetic hard disk sample. Stripes can be seen from both vertical and lateral signals, with the upwards and downwards directions being clearly visible in the V-MFM. Different between the V-MFM and L-MFM, however, is that both the sign and magnitude of the phase shift is opposite between the images. Below each image, in Fig.~\ref{fig:linescans}(c) are averaged linescans of the same area along the long stripe axis, to form horizontal line scans through Fig.~\ref{fig:linescans}(a) and (b), where this opposite phase shift sign is again apparent, with peaks in the V-MFM corresponding to valleys in the L-MFM and vice versa. 

To first address the reason for this opposite phase shift, one can in the case of long stripe domains assume without loss of generality that all field lines are lying in the x-z-plane, i.e. $H_y=0$. Consequently 
\begin{equation}\label{eq:H_y_derivatives_zero}
    \frac{\partial H_y}{\partial z} = 0 \quad\text{and}\quad \frac{\partial H_y}{\partial x} = 0.
\end{equation}
And if one considers that $\vec{\nabla} \times \vec{H} = 0$, it follows that
\begin{equation}\label{eq:H_z_derivative_zero}
    \frac{\partial H_z}{\partial y} = 0.
\end{equation}
From Maxwell's equations it follows that for every component $H_i$ of the stray field above the sample:
\begin{equation}\label{eq:laplaceH_zero_componentwise}
    \Delta H_i = \left(\frac{\partial^2}{\partial x^2} + \frac{\partial^2}{\partial y^2} + \frac{\partial^2}{\partial z^2}\right) H_i = 0.
\end{equation}
Qualitatively summing the V-MFM and L-MFM phase shifts and neglecting their different proportionality constants, one finds then:
\begin{equation}\label{eq:invertedContrast_derivation}
\begin{split}
    \left(\Delta \phi_{\text{V-MFM}} + \Delta \phi_{\text{L-MFM}}\right) &=  \frac{\partial^2}{\partial z^2}H_z + \frac{\partial^2}{\partial x^2}H_z \\
    &=  -\frac{\partial^2}{\partial y^2}H_z = 0,
\end{split}
\end{equation}
which qualitatively explains why the phase shift between V-MFM and L-MFM is opposite in this case.\\

%the stripes are pres shows linescans performed along the red line highlighted in \ref{fig:intro}(d). 
%In figures (a) through (d), the cantilever is oriented parallel to the domains, consequently the in-plane motion of the tip is oriented orthogonal to the domains. Good contrast can be seen for the V-MFM signal in \ref{fig:linescans}(a) as well as for the in-plane  L-MFM signal in \ref{fig:linescans}(b).
%Theory as to why goes here, and discuss Fig. \ref{fig:linescans}(d) here also.

It is worth noting that the maximum difference in phase is much greater for the V-MFM signal than the L-MFM. 
The V-MFM and L-MFM response was then qualitatively calculated by numerically computing the derivatives discussed in the ``theoretical considerations'' section. In Fig.~\ref{fig:linescans}(d), the qualitative numerical V-MFM and  L-MFM response is plotted, showing good agreement with the experimental data. Namely, the different peak heights between narrow and wide domains are accurately reproduced, as well as the indent in the wide peaks and the inverted contrast. \\

\begin{figure}[ht]
\includegraphics[]{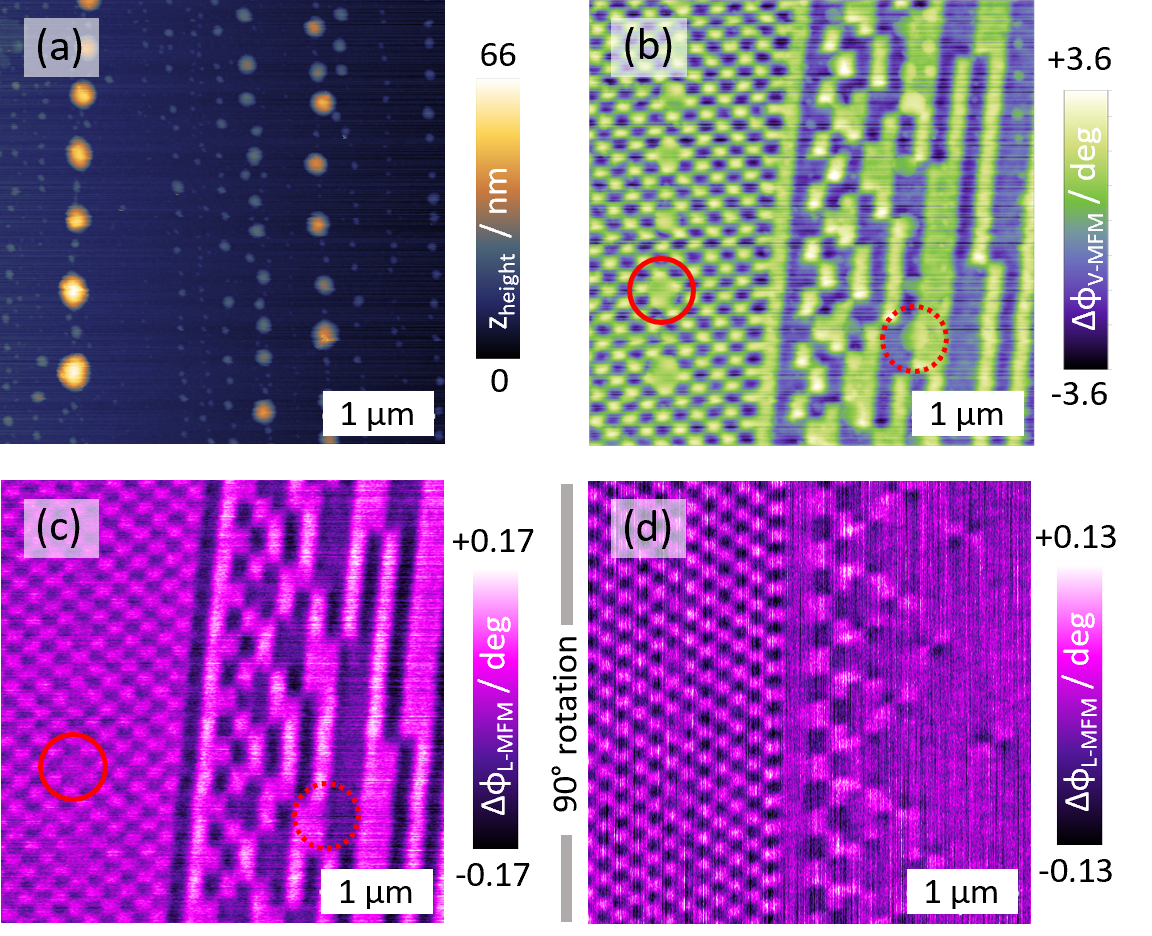}
\caption{\label{fig:2D_image}(a) shows the topography and (b) the V-MFM phase image of the hard disk. (c) shows the L-MFM phase image, with the in-plane oscillation direction oriented horizontally. For the L-MFM image in (d), the in-plane oscillation direction has been switched to be vertical.}
\end{figure}

90°-rotation of the sample (which effectively can be thought of as a rotation of the cantilever and torsional axis with respect to the sample's stripes) can be seen in  Fig.~\ref{fig:linescans}(e-h). Whereas before, the long axis of the cantilever is oriented parallel to the domains, now the cantilever is oriented orthogonal to the domain walls, with the lateral motion of the tip acting across, not along stray field lines. 
The V-MFM signal still retains similar phase shifts as acquired before, as seen in Fig.~\ref{fig:linescans}(e). The L-MFM signal in Fig.~\ref{fig:linescans}(f), however, now loses all its contrast, as also seen in the averaged linescan plot in Fig.~\ref{fig:linescans}(g). 
Intuitively, this can be understood by considering Eq. \ref{eq:LMFMPhase}. In the case of the long stripes with the assumption that the domains are magnetized entirely out-of-plane, and neglecting magnetic moment rotation within domain walls, the field lines are expected to lie in a plane orthogonal to the surface and orthogonal to the domains [see Fig.~\ref{fig:intro}(e)]. Moreover, due to the symmetry of the stripes, the field lines are expected not to change along the direction of the domains, and therefore, the derivative of any field component along this direction is expected to vanish. Since the  L-MFM signal is sensitive to exactly this derivative, this explains the loss of contrast in this case. The numerical results in Fig.~\ref{fig:linescans}(h) also show the loss of contrast for the  L-MFM signal. The difference between the experimental and numerical V-MFM signal can be explained by the tilt of the tip, which is not included in the model, but can influence the V-MFM signal \cite{rugar_magnetic_1990}. \\

%In~\ref{fig:linescans}(e-h), the cantilever is rotated by 90 degrees and now oriented perpendicular to the domains, the in-plane oscillation is therefore now oriented parallel to the domains. 
\subsubsection{Topographic interference}
Besides looking at only vertically aligned stripes, the new method was also tested on a 2D area of both the previously imaged stripes and the more complex chessboard array of up and down domains, with the results shown in Fig.~\ref{fig:2D_image}. The stray field emanating from these objects is naturally less uniaxial and thus no change in contrast when rotating the torsional axis of the chessboard domains is expected. A topography image acquired simultaneously on the first of the two dual pass images can be seen in Fig.~\ref{fig:2D_image}(a). In Fig.~\ref{fig:2D_image}(b-c), the in-plane oscillation direction is oriented horizontally, the V-MFM image is shown in Fig.~\ref{fig:2D_image}(b), the corresponding L-MFM image, recorded during the same pass, is pictured in Fig.~\ref{fig:2D_image}(c). Besides the aforementioned inverted contrast and lower signal amplitude, here the topography crosstalk visible in the V-MFM image, seems to be somewhat suppressed in the L-MFM image, particularly in the chessboard area where topographic cross talk in the V-MFM is not present in the L-MFM (see red circles). The likely cause of the lack of crosstalk is due to the fact that the torsional mode in L-MFM has only lateral contributions, unlike the inherently vertical motion that contributes to the V-MFM signal. This presents itself a unique method to determine if phase shifts in the V-MFM are caused by topographic interference or are genuinely magnetic, if they do not appear in the L-MFM, however it is by no means a universal method given the dependency of L-MFM contrast on the torsional axis being perpendicular to any in-plane stray fields. 

When again rotating the cantilever (i.e. the sample by 90°), and therefore the in-plane oscillation direction to be vertical now, and the L-MFM image in Fig.~\ref{fig:2D_image}(d) differs drastically from before. In areas where in the V-MFM image vertically oriented domains in stripe shape are visible (mostly to the right), the L-MFM image has no contrast. Domain walls, that are oriented parallel to the in-plane oscillation (i.e. vertical) are no longer visible, as expected from the linescan results. This again nicely demonstrates that the L-MFM signal is indeed sensitive to in-plane components of the stray field. \\
The most likely reason why the V-MFM and L-MFM signal could be recorded during the same pass without for example the in-plane motion affecting the spatial resolution of the V-MFM signal, is the very small mechanical amplitude of the in-plane oscillation relative to the out-of-plane oscillation amplitude \cite{kawai_ultrasensitive_2010}. Measurements were performed removing the V-MFM excitation at $\sim 70 \text{kHz}$ during the second pass with no change in the L-MFM signal found. 

\section{\label{sec:Conclusions}Conclusions}   
A bespoke, home-made cantilever holder with  split electrodes was found to reliably excite both the flexural and torsional mode of an MFM cantilever simultaneously, resulting in a mixed in- and out-of-plane oscillation of the tip. Theoretical considerations of the tip oscillation directions predicted a sensitivity of the lateral-MFM to stray fields with in-plane components along the same axis as the in-plane tip oscillation. When applied to a uni-axially magnetized hard disk sample, the difference in phase of the in-plane motion could be used to image magnetic domains of a hard disk, when the torsional axis of the cantilever was aligned such that the tip's in-plane oscillation direction lay along the same direction as the in-plane components of the protruding magnetic stray field. Similarly, when the cantilever and sample were rotated by 90$^o$ with respect to each other, the traditional V-MFM imaging provided an equivalent image of the domains, however the L-MFM imaging yielded no contrast, as the tip oscillation direction was then aligned with the domain walls, and thus detected no stray field components along it's axis. Examination of V- and L-MFM images showed a lower topographic cross talk in the case of L-MFM images, likely due to the lack of vertical component in the lateral tip oscillation direction. No cross-talk between V- and L-MFM imaging themselves was found, allowing for the simultaneous imaging of both to acquire a more complete understanding of the stray fields emanating from a sample's surface, more specifically the confirmation or absence of in-plane components within the stray fields.  

\section{Data Availability Statement}
The data that support the findings of this study are available from the corresponding author upon reasonable request.

\section{Conflicts of Interest}
The authors declare no conflicts of interest. 

\section{Acknowledgements}
The authors thank Ralf Raupach for his technical support. This work was in part supported by the Deutsche Forschungsgemeinschaft funder Grant No. SFB 1143 (Project ID No. 247310070) and the cluster of excellence ct.qmat (EXC 2147, Project ID No. 390858490).

\bibliography{LMFM.bib}% Produces the bibliography via BibTeX.

\end{document}